\documentclass[12pt,letterpaper]{article}
\usepackage[pdftex]{graphicx,color}
\usepackage{hyperref}
\input{psfig}
\usepackage{amsmath,amssymb}
\usepackage[dvips,letterpaper,text={6.5in,9in}]{geometry}
\usepackage{fancyhdr}
\usepackage{verbatim}

\newcommand\ltap{\
  \raise.3ex\hbox{$<$\kern-.75em\lower1ex\hbox{$\sim$}}\ }
\newcommand\gtap{\
  \raise.3ex\hbox{$>$\kern-.75em\lower1ex\hbox{$\sim$}}\ }

\newcommand\simge{\mathrel{%
   \rlap{\raise 0.511ex \hbox{$>$}}{\lower 0.511ex \hbox{$\sim$}}}}
\newcommand\simle{\mathrel{
   \rlap{\raise 0.511ex \hbox{$<$}}{\lower 0.511ex \hbox{$\sim$}}}}

\newcommand{\slashchar}[1]%
        {\kern .25em\raise.18ex\hbox{$/$}\kern-.75em #1}
\def\lsim{\mathrel{\raise.3ex\hbox{$<$\kern-.75em\lower1ex\hbox{$\sim$}}}}
\def\gsim{\mathrel{\raise.3ex\hbox{$>$\kern-.75em\lower1ex\hbox{$\sim$}}}}
\newcommand{\bs}{\boldsymbol}

\newcommand\CL{{\cal L}}

\newcommand\be{\begin{equation}}
\newcommand\ee{\end{equation}}
\newcommand\bea{\begin{eqnarray}}
\newcommand\eea{\end{eqnarray}}
\newcommand\ba{\begin{array}}
\newcommand\ea{\end{array}}

\newcommand\ra{\rightarrow}

\newcommand\gev{{\rm GeV}}

\newcommand\pb{{\rm pb}}

\newcommand\fb{{\rm fb}}
\newcommand\ifb{{\rm fb}^{-1}}

\newcommand\etmiss{\slashchar{E}_T}

\newcommand\ellm{\ell^-}
\newcommand\ellpm{\ell^\pm}
\newcommand\ellp{\ell^+}

\newcommand\Ntc{N_{TC}}

\newcommand\tom{\omega_{T}}
\newcommand\tro{\rho_{T}}

\newcommand\ta{a_T}

\newcommand\tropm{\rho_{T}^\pm}

\newcommand\tpi{\pi_T}
\newcommand\tpipm{\pi_T^\pm}

\newcommand\tpiz{\pi_T^0}

\newcommand\Mjj{M_{jj}}

\newcommand\MWjj{M_{Wjj}}
\newcommand\MZjj{M_{Zjj}}

\begin{document}

\title{
\vskip -15mm
\begin{flushright}
 \vskip -15mm
 {\small FERMILAB-Pub-11-335-T\\
 }
 \vskip 5mm
 \end{flushright}
{\Large{\bf Testing CDF's Dijet Excess \\and Technicolor at the LHC}}\\
} \author{
  {\large Estia Eichten$^{1}$\thanks{eichten@fnal.gov} ,\,
  Kenneth Lane$^{2}$\thanks{lane@physics.bu.edu} \, and
  Adam Martin$^{1}$\thanks{aomartin@fnal.gov}}\\
{\large {$^{1}$}Theoretical Physics Group, Fermi National Accelerator Laboratory}\\
{\large P.O. Box 500, Batavia, Illinois 60510}\\
{\large $^{2}$Department of Physics, Boston University}\\
{\large 590 Commonwealth Avenue, Boston, Massachusetts 02215}\\
}
\maketitle

\begin{abstract}

  Under the assumption that the dijet excess seen by the CDF Collaboration
  near $150\,\gev$ in $Wjj$ production is due to the lightest technipion of
  the low-scale technicolor process $\tro \ra W\tpi$, we study its
  observability in LHC detectors with $1-5\,\ifb$ of data. We find that
  cuts similar to those employed by CDF are unlikely to confirm its
  signal. We propose cuts tailored to the LSTC hypothesis and its backgrounds
  at the LHC that can reveal $\tpi \ra jj$. We also stress the importance at
  the LHC of the isospin-related channel $\tropm \ra Z\tpipm \ra
  \ellp\ellm\,jj$ and the all-lepton mode $\tropm \ra WZ \ra
  \ellp\ellm\ellpm\nu_\ell$.

\end{abstract}



 




\newpage

\section*{1. Introduction}

The CDF Collaboration recently reported evidence for a resonant excess near
$150\,\gev$ in the dijet-mass spectrum, $\Mjj$, of $Wjj$
production~\cite{Aaltonen:2011mk}. For an integrated luminosity of
$4.3\,\ifb$, CDF fit the excess to a simple Gaussian with $\sigma_{\rm
  resolution} = 14.3\,\gev$ and determined its significance to be
$3.2\,\sigma$ and its cross section to be ``of order $4\,\pb$''. CDF has
updated this paper using $\int \CL dt = 7.3\,\ifb$, and the significance of
the dijet excess is now $4.1\,\sigma$~\cite{CDFnew}. The D\O\ Collaboration,
on the other hand, has analyzed $4.3\,\ifb$ and reported no excess. A
$4\,\pb$ cross section is rejected at the level of $4.3\,\sigma$, while the
95\% confidence level upper limit on the cross section is
$1.9\,\pb$~\cite{Abazov:2011af}. 

In Ref.~\cite{Eichten:2011sh} we proposed a low-scale technicolor (LSTC)
explanation for CDF's dijet excess: A technirho ($\tro^{\pm,0}$) of mass
$M_{\tro} = 290\,\gev$ is produced as a very narrow $s$-channel resonance in
$\bar q q$ annihilation and decays into a technipion ($\tpi^{0,\pm}$) with
$M_{\tpi} = 160\,\gev$ plus a $W$-boson which is mostly longitudinally
polarized.\footnote{Other relevant LSTC masses are $M_{\tom} = M_{\tro}$;
  $M_{\ta} = 1.1 M_{\tro} = 320\,\gev$; and $M_{V_i,A_i}$ which appear in
  dimension-five operators for $V_T$ decays to transverse EW boson; we take
  them equal to $M_{\tro}$. Other LSTC parameters are $\sin\chi = 1/3$, $Q_U
  = Q_D + 1 = 1$, and $\Ntc = 4$.} Using the LSTC model implemented in {\sc
  Pythia}~\cite{Lane:2002sm,Eichten:2007sx,Sjostrand:2006za}, we found
$\sigma(\bar pp \ra \tro \ra W\tpi \ra Wjj) = 2.4\,\pb$. We closely matched
CDF's dijet mass distribution for the signal and background. Motivated by the
peculiar kinematics of $\tro$ production at the Tevatron and $\tro \ra W\tpi$
decay, we also suggested cuts intended to enhance the $\tpi$ signal's
significance and make $\tro \ra Wjj$ visible. Several distributions ---
$p_T(jj)$, $\Delta\phi(jj)$, $\Delta R(jj)$ and $M_{Wjj}$ --- presented by
CDF in Ref.~\cite{CDFnew} fit the expectations of the LSTC model quite
well. This will be elaborated upon in an upcoming publication.

In this note we present the results of simulations of $\tro \ra W\tpi \ra
Wjj$ at the LHC. We predict that the cross section there is $8.0\,\pb$. We
find that the cuts employed by CDF in Refs.~\cite{Aaltonen:2011mk,CDFnew} appear to be
insufficient to extract the $\tpi \ra jj$ signal from the background, even
for a data sample of $\sim 5\,\ifb$. We also find that, while cuts similar to
the ones we proposed in Ref.~\cite{Eichten:2011sh} significantly enhance the
signal-to-background, they cause the background to peak very near the
signals themselves. We therefore propose qualitatively different ones
that should give more isolated, observable $\tpi$ and $\tro$ signals for at
most a few~$\ifb$. The selections we consider are specific to our $\tro$
explanation of the CDF excess and may not be useful for testing other
proposals --- which generally do not share the peculiar kinematics of ours
(for a sampling of other proposed explanations of CDF's dijet excess, see
Refs.~\cite{Kilic:2011vn, Buckley:2011ly, Hewett:2011fk, Harnik:2011mv,
  Cao:2011qf, Dobrescu:2011fk, Nelson:2011fk, Yu:2011ve, Cheung:2011uq}).

In Ref.~\cite{Eichten:2011sh} we mentioned other processes that can be sought
at the Tevatron and LHC and which should be seen soon if the CDF signal is
real and has the LSTC origin we proposed. We highlight two of these,
$\tropm \ra Z\tpipm$ and $W^\pm Z$, at the end of this note.

\section*{2. Simulations of the CDF Signal at the LHC}

\begin{figure}[!t]
 \begin{center}
\includegraphics[width=6.50in, height=3.15in]{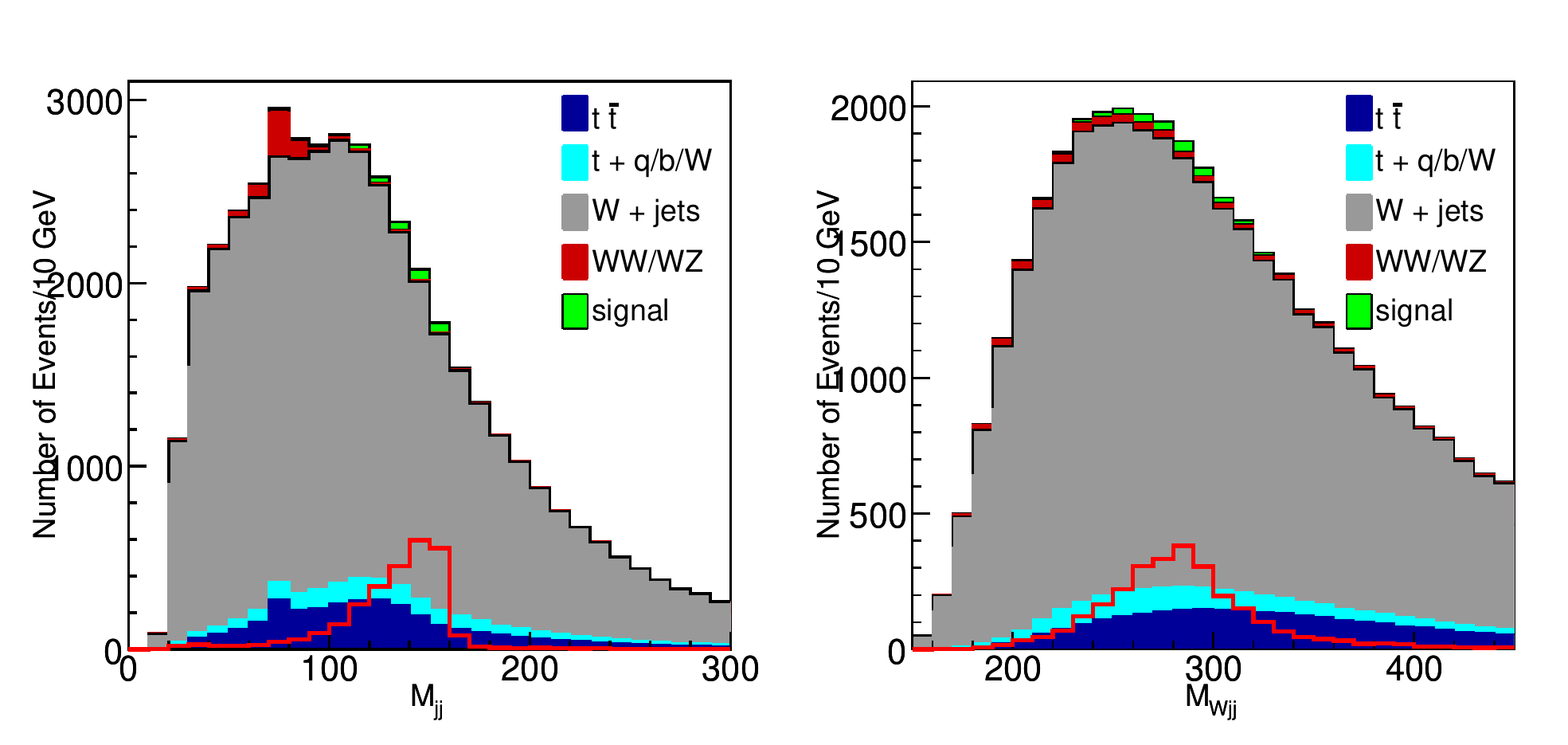}
\caption{The $\Mjj$ and $\MWjj$ distributions of $\tro \ra W\tpi \ra
  \ell\nu_\ell jj$ and its backgrounds at the LHC for $\int \CL dt =
  1\,\ifb$. CDF-like cuts as described in the text are employed. The
  important backgrounds are indicated and the $\tpi$ and $\tro$ signals
  $\times 10$ are shown as the thin red-lined histograms.
\label{fig:tclhc7_cuts0}}
 \end{center}
 \end{figure}
\begin{figure}[!h]
 \begin{center}
\includegraphics[width=6.50in, height=3.15in]{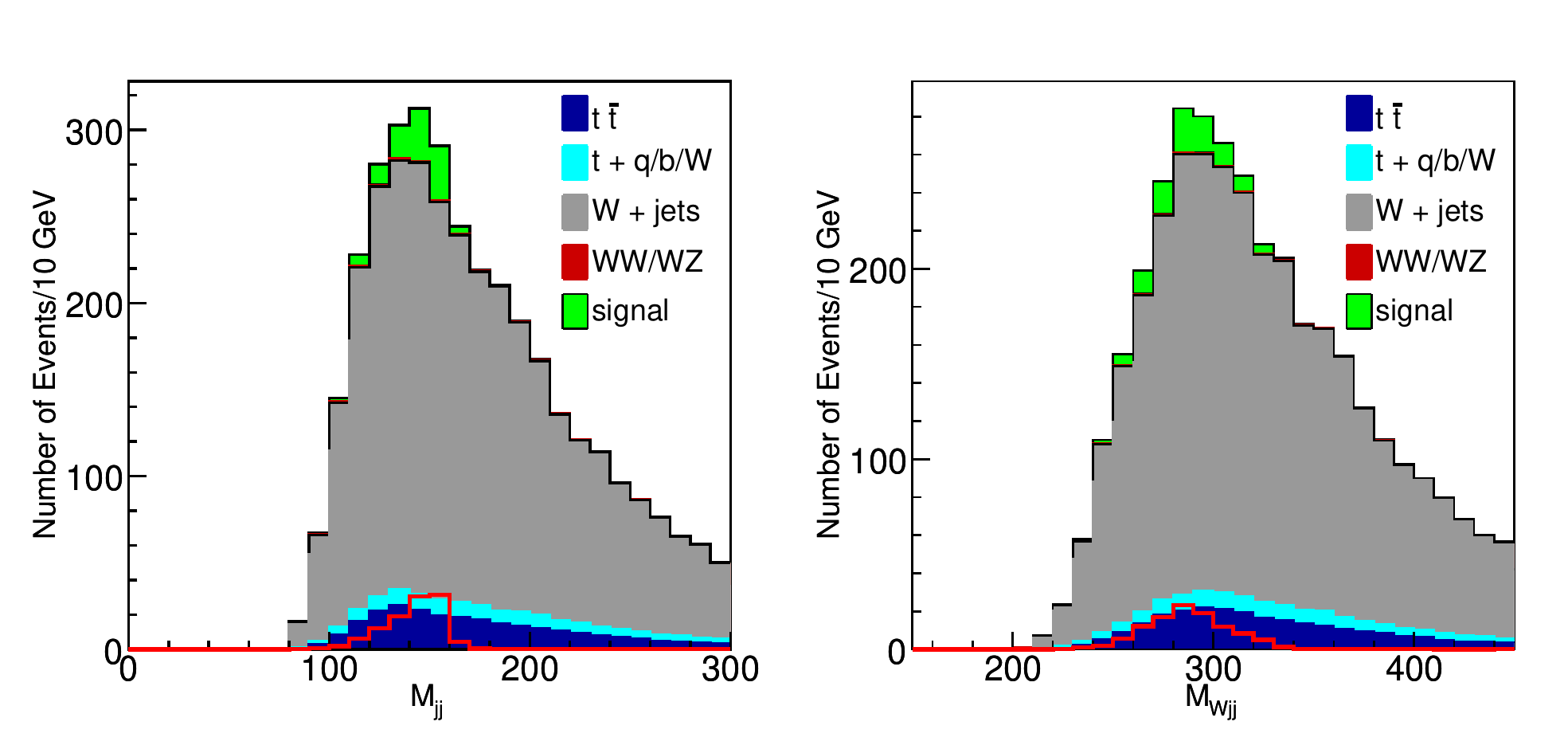}
\caption{The $\Mjj$ and $\MWjj$ distributions of $\tro \ra W\tpi \ra
  \ell\nu_\ell jj$ and its backgrounds at the LHC for $\int \CL dt =
  1\,\ifb$. Augmented CDF-like cuts, described in the text and similar to ones
  proposed in Ref.~\cite{Eichten:2011sh}, are employed. They result in
  enhanced $\tpi$ and $\tro$ signals appearing at the peaks of their
  backgrounds. The unscaled $\tpi$ and $\tro$ signals are also shown as thin
  red-lined histograms.
  \label{fig:tclhc7_cuts4}}
 \end{center}
 \end{figure}

 The obvious place to start is with the cuts employed by
 CDF~\cite{Aaltonen:2011mk}.\footnote{The CDF cuts are: exactly one lepton,
   $\ell = e,\mu$, with $p_T > 20\,\gev$ and $|\eta| < 1.0$; exactly two jets
   with $p_T > 30\,\gev$ and $|\eta| < 2.4$; $\Delta R(\ell, j) > 0.52$;
   $p_T(jj) > 40\,\gev$; $\etmiss > 25\,\gev$; $M_T(W) > 30\,\gev$;
   $|\Delta\eta(jj)| < 2.5$; $|\Delta\phi(\bs{\etmiss},j)| > 0.4$.} However,
 for $\int\CL dt = {\rm few}\,\ifb$, we believe this will be fruitless.
 Fig.~\ref{fig:tclhc7_cuts0} shows the $\Mjj$ and $\MWjj$ distributions for
 $1\,\ifb$ with CDF cuts except that we require that leptons have $p_T >
 30\,\gev$ and $|\eta| < 2.5$, reflecting the greater acceptance of the LHC
 detectors.\footnote{Backgrounds were generated at matrix-element level using
   ALPGENv213~\cite{Mangano:2002ea}, then passed to {\sc Pythia}v6.4 for
   showering and hadronization. We use CTEQ6L1 parton distribution functions
   and a factorization/renormalization scale of $\mu = 2 M_W$ throughout. For
   the dominant $W+$jets background we generate $W+2j$ (excl.) plus $W+3j$
   (inc.)  samples, matched using the MLM procedure~\cite{MLM} (patron level
   cuts are imposed to ensure that $W+0, 1$ jet events cannot
   contribute). After matching, the overall normalization is scaled to the
   NLO $W+jj$ value, calculated with MCFMv6~\cite{Campbell:2011bn}. After
   passing through {\sc Pythia}, final state particles are combined into
   $(\eta, \phi)$ cells of size $0.1\times 0.1$, with the energy of each cell
   rescaled to make it massless. Isolated photons and leptons ($e,\mu$) are
   removed, and all remaining cells with energy greater than $1\,\gev$ are
   clustered into jets using FastJet (anti-kT algorithm, $R =
   0.4$)~\cite{Cacciari:2005hq}.} As in Ref.~\cite{Eichten:2011sh}, we do not
 include calorimetric energy smearing, hence the narrow $W/Z \ra jj$ peak of
 diboson production near $80\,\gev$.  This simplification does not affect our
 $\tpi \ra jj$ mass resolution which is due mainly to jet definition. The
 background under the dijet resonance in Fig.~\ref{fig:tclhc7_cuts0} is a
 factor of 5--6 times greater than at the Tevatron; see Fig.~1 in
 Refs.~\cite{Aaltonen:2011mk,Eichten:2011sh}.  Counting events in the four
 bins from $\Mjj = 120$ to $160\,\gev$, we obtain $S/\sqrt{B} = 2.10$ and
 $S/B = 0.023$. Given this and the shape of the signal and background, it is
 doubtful that CDF-like cuts alone could provide confirmation of its dijet
 signal for even $5\,\ifb$ of data.

\begin{figure}[!t]
 \begin{center}
\includegraphics[width=6.50in, height=3.15in]{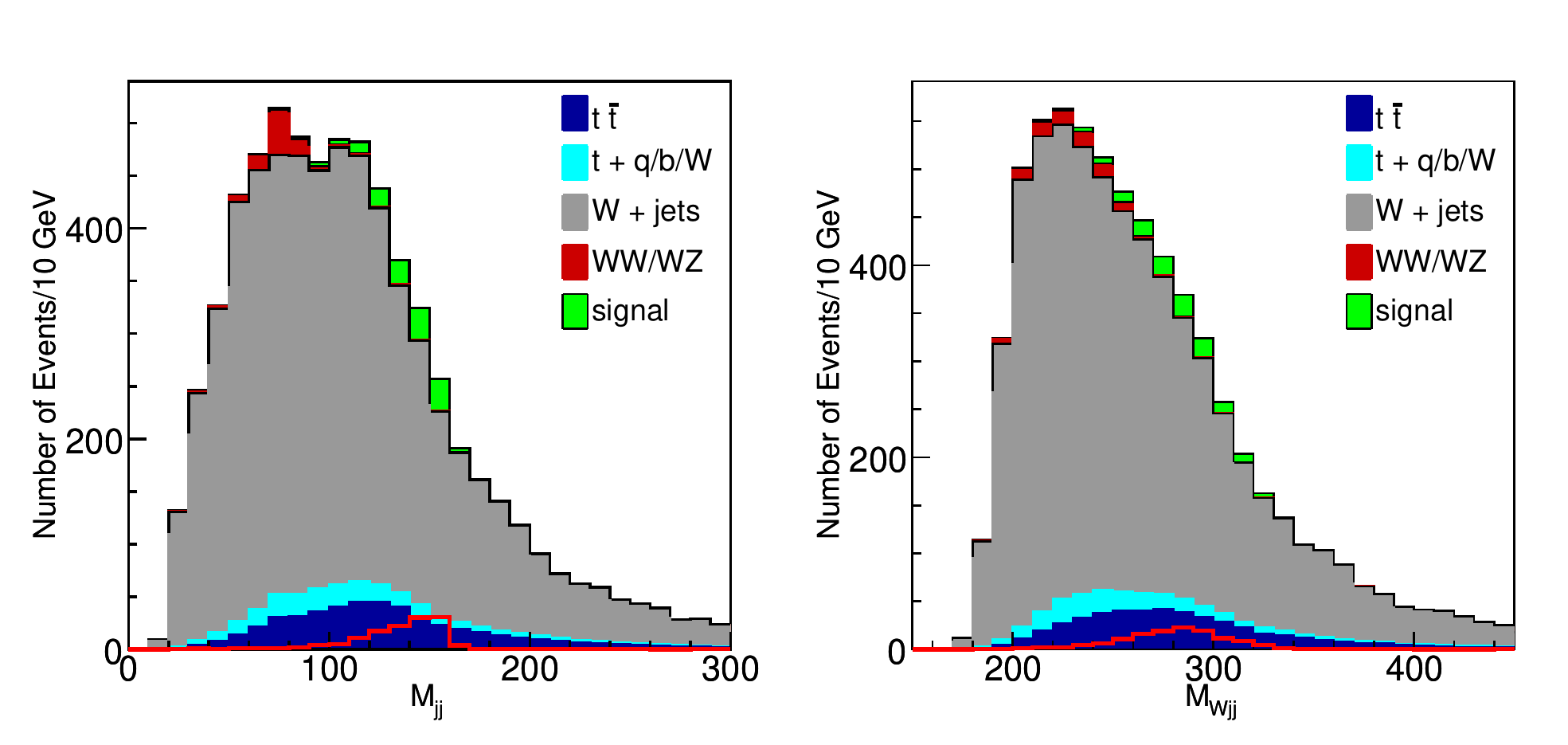}
\caption{The $\Mjj$ and $\MWjj$ distributions of $\tro \ra W\tpi \ra
  \ell\nu_\ell jj$ and its backgrounds at the LHC for $\int \CL dt =
  1\,\ifb$. Augmented CDF-like cuts as described in the text are employed.
  The enhanced $\tpi$ and $\tro$ signals now appear below the peaks of their
  backgrounds. The unscaled $\tpi$ and $\tro$ signals are also shown as thin
  red-lined histograms.
  \label{fig:tclhc7_shifted}}
 \end{center}
 \end{figure}
\begin{figure}[!h]
 \begin{center}
\includegraphics[width=6.50in, height=3.15in]{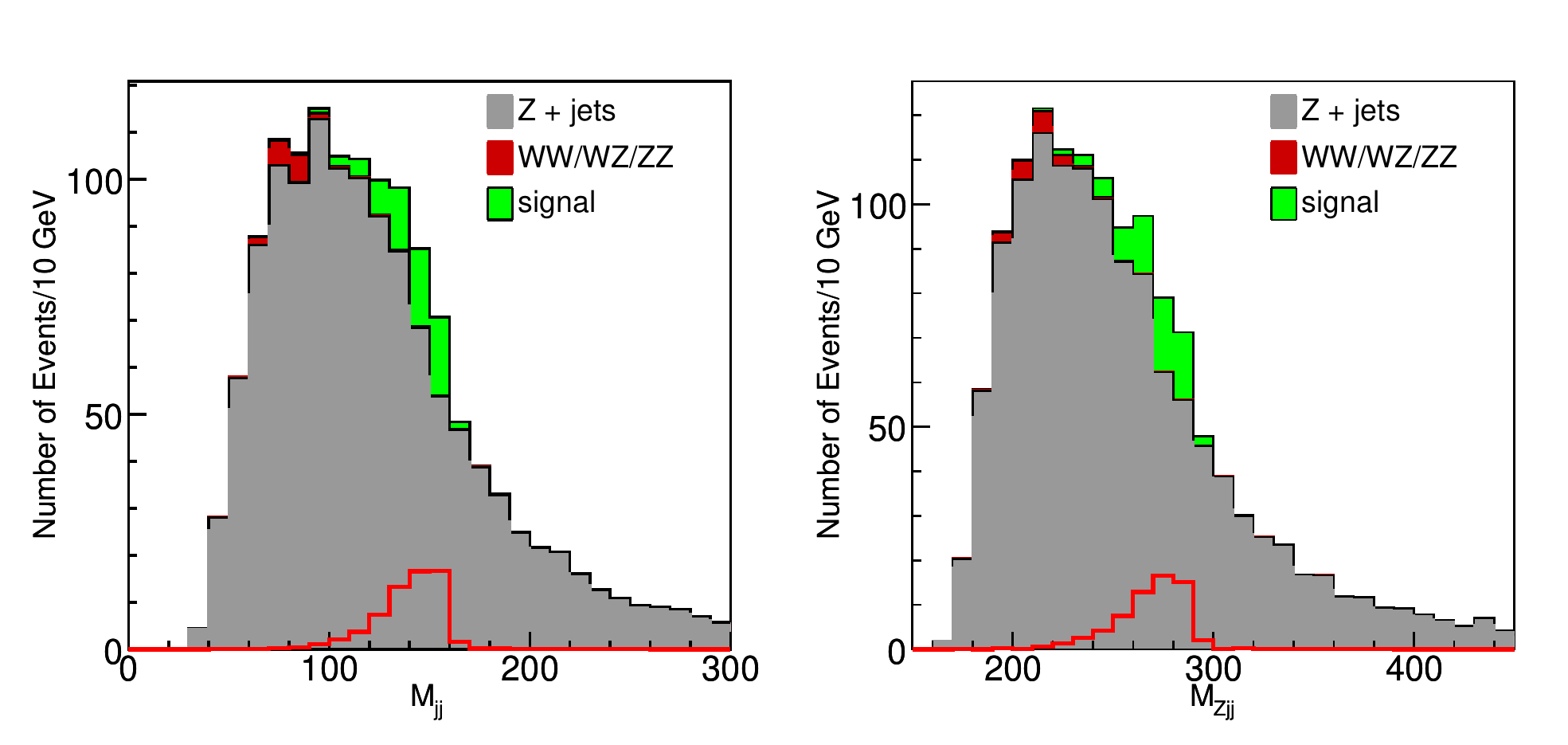}
\caption{The $\Mjj$ and $\MZjj$ distributions of $\tropm \ra Z\tpipm \ra
  \ellp\ellm jj$ and its backgrounds at the LHC for $\int \CL dt = 5\,\ifb$.
  The cuts are described in the text. The unscaled $\tpi$ and $\tro$ signals
  are also shown as thin red-lined histograms.
  \label{fig:tclhc7_zjj_cuts1}}
 \end{center}
 \end{figure}

 To improve the signal-to-background, we examined a variety of cuts motivated
 by $\tro \ra W\tpi$ kinematics and similar in character to those proposed in
 Ref.~\cite{Eichten:2011sh}. Fig.~\ref{fig:tclhc7_cuts4} was obtained
 applying the following requirements in addition to the CDF-like cuts:
 $\Delta\phi(jj) > 2.0$, $Q = \MWjj - \Mjj - M_W < 100\,\gev$, $p_T(jj) >
 60\,\gev$ and $p_T(W) > 60\,\gev$. The $\tpi$ signal now has $S/\sqrt{B} =
 2.82$ and $S/B = 0.085$. Unfortunately, as can be seen in
 Fig.~\ref{fig:tclhc7_cuts4}, these cuts cause the background to peak very
 near the dijet resonance so that the $\tpi$'s observation at the LHC would
 require not only very good understanding of the $Wjj$ backgrounds just where
 they are largest, but probably considerably more data than the $\simeq
 5\,\ifb$ expected to be collected this year.
 
 We have obtained what we believe is an acceptable separation of the
 background peak from the $\tpi$ signal with the following cuts: $p_T(j_1) >
 40\,\gev$ while $p_T(j2) > 30\,\gev$, $p_T(jj) > 45\,\gev$, $p_T(W) >
 60\,\gev$, $\Delta\eta(jj) < 1.2$ (this was 2.5 in
 Refs.~\cite{Aaltonen:2011mk,Eichten:2011sh}) and $Q > 20\,\gev$. The results
 are shown in Fig.~\ref{fig:tclhc7_shifted}. Counting events gives
 $S/\sqrt{B} = 2.80$ and $S/B = 0.078$ for the $\tpi \ra jj$ signal and $\int
 \CL dt = 1\,\ifb$. A valuable feature of this selection is the diboson
 production $W/Z \ra jj$ peak near $80\,\gev$. It allows self-calibration of
 the background normalization at its peak. With proper cuts on only a few
 $\ifb$ of data, therefore, the LHC experiments should be able to confirm or
 exclude the $\tpi$ signal. The $\tro \ra Wjj$ signal in the interval $260 <
 \MWjj < 300\,\gev$ in Fig.~\ref{fig:tclhc7_shifted} has $S/\sqrt{B} = 2.50$
 and $S/B = 0.089$ for $1\,\ifb$. It should be observable with $\sim
 5\,\ifb$.


\section*{3. The $\tropm \ra Z\tpipm$ and $W^\pm Z$ Modes}

An important confirmation of the $\tro \ra W\tpi \ra \ell\nu_\ell jj$ signal
(albeit, one not free of all $Wjj$ background issues) is observation of its
isospin partner, $\tropm \ra Z\tpipm \ra \ellp\ellm jj$. Because of the
limited phase space in these decays, the {\sc Pythia} cross section at the
LHC for $\tropm \ra Z \tpipm$ is only $2.36\,\pb$ compared to $3.44\,\pb$ for
$\tropm \ra W^\pm \tpiz$~\footnote{This assumes $B(\tpipm\ra \bar
  q'q)/B(\tpiz \ra \bar qq) \simeq 1$. The cross section ratio agrees well
  with $p_Z^3/p_W^3 = 0.69$.} and $7.90\,\pb$ for both $\tro \ra W\tpi$
channels. The branching ratio for $Z \ra e^+e^-,\, \mu^+\mu^-$ reduces this
to $165\,\fb$, 10\% of the $\tro \ra \ell\nu_\ell jj$ rate. Thus, for a
similar ratio of backgrounds, we expect that $\sim 10$ times the luminosity
needed for the $\tro \ra W\tpi$ signal would be required for the same
sensitivity. Actually, because the $Zjj$ background is less than 10\% of the
$Wjj$ background, the situation is better than this and just $5\,\ifb$ are
needed to give $S/\sqrt{B} =3.12$ and $S/B = 0.18$; see
Fig.~\ref{fig:tclhc7_zjj_cuts1}. The cuts used there are: two electrons or
muons of opposite charge with $p_T > 30\,\gev$ and $|\eta| < 2.5$, exactly
two jets with $p_T > 30\,\gev$ and $|\eta| < 2.5$, $p_T(jj) > 40\,\gev$,
$p_T(Z) > 50\,\gev$, $\Delta\eta(jj) < 1.75$, and $Q < 60\,\gev$.

Finally, the mode $\tropm \ra W^\pm Z \ra \ellpm\nu_\ell \ellp\ellm$ is
another important check on the LSTC hypothesis~\cite{Brooijmans:2010tn}. We
expect $\sigma(\tropm \ra W^\pm Z)/\sigma(\tropm \ra W^\pm \tpiz) =
(p(Z)/p(\tpi))^3 \tan^2\chi$. The {\sc Pythia} rate agrees with this
estimate; for $\sin\chi = 1/3$ and our input masses, $\sigma(pp \ra \tropm
\ra \ellpm\nu_\ell\ellp\ellm) = 25\,\fb$ at the LHC. This should be
observable with $\int \CL dt \simeq 5\,\ifb$.

\section*{Acknowledgments} 

We are grateful to K.~Black, T.~Bose, P.~Catastini, V.~Cavaliere,
C.~Fantasia, E.~Pilon, for valuable conversations and advice. This work was
supported by Fermilab operated by Fermi Research Alliance, LLC,
U.S.~Department of Energy Contract~DE-AC02-07CH11359 (EE and AM) and in part
by the U.S.~Department of Energy under Grant~DE-FG02-91ER40676~(KL). KL's
research was also supported in part by Laboratoire d'Annecy-le-Vieux de
Physique Theorique (LAPTh) and the CERN Theory Group and he thanks LAPTh and
CERN for their hospitality.

\vfil\eject

\bibliography{TC_at_LHC}
\bibliographystyle{utcaps}
\end{document}